\newif\ifdouble
\doubletrue

\newif\ifFullVersion
\FullVersiontrue  

\ifdouble
\documentclass[9pt, journal]{IEEEtran}
\else
\documentclass[12pt, draftclsnofoot, onecolumn]{IEEEtran}
\fi
\usepackage{amsthm}
\usepackage{latexsym}
\usepackage{amssymb}
\usepackage{graphicx}
\usepackage{subcaption}
\usepackage{amsmath}
\usepackage{color}
\usepackage{cite}
\usepackage[bookmarks,colorlinks]{hyperref}
\usepackage{cleveref}
\usepackage{acronym}
\usepackage[english]{babel}
\usepackage{algpseudocode}
\usepackage[Algorithmus]{algorithm}
\algnewcommand{\IIf}[1]{\State\algorithmicif\ #1\ \algorithmicthen}
\algnewcommand{\EndIIf}{\unskip\ \algorithmicend\ \algorithmicif}
\newcommand{\rev}[1]{\textcolor[rgb]{0.00,0.00,0.00}{#1}}

\theoremstyle{plain}
\newtheorem{theorem}{Theorem}

\theoremstyle{definition}

\theoremstyle{remark}

\newcommand{\off}[1]{}

\ifCLASSINFOpdf
\else
\fi

\begin{document}

\title{Distributed Quantization for Sparse Time Sequences}
\markboth{}{}
\author{%
	\IEEEauthorblockN{Alejandro Cohen\IEEEauthorrefmark{1},
	Nir Shlezinger\IEEEauthorrefmark{2},
	Salman Salamatian\IEEEauthorrefmark{1},
	Yonina C. Eldar\IEEEauthorrefmark{2},
	and Muriel M\'edard\IEEEauthorrefmark{1}}\\
	\IEEEauthorblockA{\IEEEauthorrefmark{1}%
		Research Laboratory of Electronics, MIT, Cambridge, MA, USA,
		\{cohenale, salmansa, medard\}@mit.edu}\\
	\IEEEauthorblockA{\IEEEauthorrefmark{2}%
		Math and CS, Weizmann Institute of Science, Rehovot, Israel,
		\{nir.shlezinger, yonina.eldar\} @weizmann.ac.il}
	\vspace{-10mm}
		\thanks{
		This project received funding from the Benoziyo Endowment Fund for the Advancement of Science, the	Estate of Olga Klein - Astrachan, the European Union’s Horizon 2020 research and innovation program under grant No. 646804-ERC-COG-BNYQ, and from the Israel Science Foundation under grant No. 0100101.
	}
	}
\maketitle
\pagestyle{empty}
\thispagestyle{empty}
\definecolor{NewColor}{rgb}{0,0,0.4}

\newcommand{\myVec}[1]{{\boldsymbol{#1}}}
\newcommand{\myMat}[1]{{\boldsymbol{#1}}}
\newcommand{\mySet}[1]{\mathcal{#1}}

\newcommand{\myDetVec}[1]{\myVec{\lowercase{#1}}}
\newcommand{\myRandVec}[1]{\myVec{\lowercase{#1}}}
\newcommand{\myDetMat}[1]{\myMat{\uppercase{#1}}}
\newcommand{\myRandMat}[1]{\myMat{\uppercase{#1}}}

\newcommand{\E}{\mathbb{E}}		 				
\newcommand{\myW}{{\myRandVec{W}}}			 		
\newcommand{\myY}{{\myRandVec{Y}}}			 		
\newcommand{\myX}{{\myRandVec{X}}}	
\newcommand{\myZ}{{\myRandVec{Z}}}		 		
\newcommand{\myV}{{\myRandVec{V}}}			 		
\newcommand{\myS}{{\myDetVec{s}}}			 		
\newcommand{\myI}{{\myDetMat{i}}}			 		
\newcommand{\myA}{{\myDetMat{a}}}
\newcommand{\myB}{{\myDetMat{b}}}
\newcommand{\myAT}{\tilde{\myA}}
\newcommand{\myBT}{\tilde{\myB}}			 					 		
\newcommand{\myAB}{\bar{\myA}}
\newcommand{\myYmat}{{\myRandMat{Y}}}			 	
\newcommand{\myYvec}{\underline{\myY}}			 	
\newcommand{\myQvec}{\underline{\myVec{q}}}			 	
\newcommand{\mySmat}{{\myMat{\Theta}}}			 	
\newcommand{\myWmat}{{\myRandMat{W}}}			 	
\newcommand{\myWvec}{\underline{\myW}}			 	
\newcommand{\Gmat}{{\myRandMat{G}}}			 		
\newcommand{\Gvec}{\underline{\myVec{g}}}			 		
\newcommand{\GmatRel}{{\myMat{g}}}			 		
\newcommand{\Hmat}{{\myRandMat{H}}}			 		
\newcommand{\Hvec}{\underline{\myVec{h}}}
\newcommand{\Dmat}{{\myDetMat{d}}}			 		
\newcommand{\Bmat}{{\myDetMat{f}}}			 		
\newcommand{\Phimat}{{\myMat{\Phi}}}			 		
\newcommand{\DmatRel}{\bar{\myDetMat{d}}}			 		
\newcommand{\BmatRel}{\bar{\myDetMat{f}}}			 		
\newcommand{\AggMat}{{\myRandMat{A}}}			 	
\newcommand{\Ymat}{\tilde{\myRandMat{Y}}}			
\newcommand{\Wmat}{\tilde{\myRandMat{W}}}			
\newcommand{\Smat}{{\myMat{\Theta}}}				
\newcommand{\myTheta}{\theta}
\newcommand{\SigW}{\sigma_W^2}						
\newcommand{\AntRatio}{\kappa}						
\newcommand{\Ncells}{n_c}							
\newcommand{\Nantennas}{\lenX}						
\newcommand{\Nusers}{\lenS}							
\newcommand{\NcellsSet}{\mySet{N}_c}				
\newcommand{\NusersSet}{\mySet{K}}				
\newcommand{\dcoeff}{d}								
\newcommand{\dcoeffRel}{d}								
\newcommand{\bcoeff}{f}								
\newcommand{\phicoeff}{\phi}								
\newcommand{\Tpilots}{\lenXtag}						
\newcommand{\TpilotsSet}{\mySet{L}}						
\newcommand{\Tdata}{\tau_d}							
\newcommand{\Tcoh}{\tau_c}							
\newcommand{\EstGmat}{\hat{\Gmat}}					
\newcommand{\ErrGmat}{\tilde{\Gmat}}				
\newcommand{\MPFunc}{\nu}								
\newcommand{\SemiCirc}{F}							
\newcommand{\UHmat}{ \myMat{M}}
\newcommand{\Sol}{s}
\newcommand{\Dist}{ \stackrel{d}{=}}
\newcommand{\AsConv}{\mathop{\longrightarrow}\limits^{\rm a.s.}}
\newcommand{\CDF}[1]{F_{#1}}
\newcommand{\Pdf}[1]{f_{ #1}}
\newcommand{\Psd}[1]{s_{#1}}
\newcommand{\Acorr}[1]{c_{#1}}
\newcommand{\PSD}[1]{\myMat{S}_{#1}}
\newcommand{\ACORR}[1]{\myMat{C}_{#1}}
\newcommand{\CorrMat}[1][ ]{\myMat{C}_{#1}}
\newcommand{\CovMat}[1]{\myMat{\Sigma}_{#1}}			
\newcommand{\CovMatExt}[1]{{\underline{\myMat{\Sigma}}}_{#1}}			
\newcommand{\maxDiag}{\sigma^2_{l}}
\newcommand{\bits}{b}
\newcommand{\SpaSize}{k}
\newcommand{\Rate}{R}
\newcommand{\Ratio}{r}
\newcommand{\AsymDist}{\mu} 
\newcommand{\lenX}{n}			 			
\newcommand{\lenT}{T}			 			
\newcommand{\lenZ}{p}			 			
\newcommand{\lenZT}{\tilde{\lenZ}}			 			
\newcommand{\lenZn}{m_p}
\newcommand{\lenZq}{m_q}
\newcommand{\lenSset}{\mySet{K}}			 			
\newcommand{\lenXset}{\mySet{N}}	
\newcommand{\lenTset}{\mySet{T}}
\newcommand{\Quan}[2]{Q_{{#1}}^{{#2}}}
\newcommand{\LmmseMat}{\myMat{\Gamma}}
\newcommand{\LmmseMatT}{\tilde{\LmmseMat}}
\newcommand{\EmpSet}{\varnothing}
\newcommand{\DynRange}{\gamma}
\newcommand{\DynInt}[1][ ]{\Delta_{#1}}
\newcommand{\TilM}[1][ ]{\tilde{M}_{#1}}
\newcommand{\MyKappa}[1][]{\kappa_{#1}}
\newcommand{\Qnoise}{\myVec{e}}
\newcommand{\Wlevel}{\zeta}
\newcommand{\myEta}{\eta}
\newcommand{\DistG}{D_{\rm G}}
\newcommand{\MMSE}{^{\rm MMSE}}
\newcommand{\Opt}{^{\rm Opt}}
\newcommand{\op}{^{\rm o}}
\newcommand{\Ign}{^{\rm Ign}}
\newcommand{\ADC}{^{\rm HL}}
\newcommand{\sADC}{^{\rm sHL}}
\newcommand{\myObs}{\underline{\myObstag}}
\newcommand{\mySOI}{\underline{\mySOItag}}
\newcommand{\mySOIEst}{\underline{\mySOIEsttag}}
\newcommand{\myQ}{\myVec{q}}
\newcommand{\lenXtag}{L}
\newcommand{\lenS}{m}
\newcommand{\myObstag}{\myVec{y}}
\newcommand{\mySOItag}{\myVec{g}}
\newcommand{\mySOIEsttag}{\tilde{\mySOItag}}
\newcommand{\LmmseMattag}{{\LmmseMat}}
\newcommand{\eig}[1]{\lambda_{#1}}			
\newcommand{\eigT}[1]{\eig{#1}}
\newcommand{\CovYtag}{\CovMat{\myY_l}}
\newcommand{\myAtag}{\myA\op}
\newcommand{\myBtag}{\myB\op}
\newcommand{\Glevel}{\varphi}
\newcommand{\GlevelT}{\tilde{\varphi}}
\newcommand{\Plevel}{\Glevel}

\newcommand{\lenL}{l}			 			
\newcommand{\myBin}{\mySet{B}}
\newcommand{\mySubBin}{\mySet{SB}}
\newcommand{\myCodeword}{\myVec{c}}
\newcommand{\myCodewordT}{\tilde{\myVec{c}}}
\newcommand{\ScaQuant}{q}

\acrodef{bs}[BS]{base station}
\acrodef{mimo}[MIMO]{multiple-input multiple-output}
\acrodef{mac}[MAC]{multiple access channel}
\acrodef{dsp}[DSP]{digital signal processor}
\acrodef{ut}[UT]{user terminal}
\acrodef{cdf}[CDF]{cumulative distribution function}
\acrodef{pdf}[PDF]{probability density function}
\acrodef{ps}[PS]{pilot sequence}
\acrodef{se}[SE]{spectral efficiency}
\acrodef{mse}[MSE]{mean-squared error}
\acrodef{adc}[ADC]{analog-to-digital convertor}
\acrodef{dtft}[DTFT]{discrete-time Fourier transform}
\acrodef{dft}[DFT]{discrete Fourier transform}
\acrodef{nb}[NB]{narrowband}
\acrodef{dt}[DT]{discrete-time}
\acrodef{ct}[CT]{continuous-time}
\acrodef{evd}[EVD]{eigenvalue decomposition}
\acrodef{svd}[SVD]{singular valued decomposition}
\acrodef{soi}[SOI]{signal of interest}
\acrodef{awgn}[AWGN]{additive white Gaussian noise}
\acrodef{wss}[WSS]{wide-sense stationary}
\acrodef{mmse}[MMSE]{minimum \ac{mse}}
\acrodef{mi}[MI]{mutual information}
\acrodef{lmmse}[LMMSE]{linear MMSE}
\acrodef{map}[MAP]{maximum a-posteriori probability}
\acrodef{ml}[ML]{maximum likelihood}
\acrodef{isi}[ISI]{intersymbol interference}
\acrodef{snr}[SNR]{signal-to-noise ratio}
\acrodef{pc}[PC]{proper-complex}
\acrodef{cs}[CS]{Compressed sensing}
\acrodef{psd}[PSD]{power spectral density}
\acrodef{ptp}[PtP]{point-to-point}
\acrodef{sinr}[SINR]{signal-to-interference-and-noise ratio}
\acrodef{pdf}[PDF]{probability density function}
\acrodef{rv}[RV]{random variable}
\acrodef{csi}[CSI]{channel state information}
\acrodef{sqrss}[DiSeQuanS]{distributed serial quantization of sparse signals}
\acrodef{qiht}[QIHT]{quantized iterative hard thresholding}
\acrodef{fista}[FISTA]{fast iterative soft thresholding algorithm}  %

\begin{abstract}
Analog signals processed in digital hardware are quantized into a discrete bit-constrained representation.  Quantization is typically carried out using analog-to-digital converters (ADCs), operating in a serial scalar manner. In some applications, a set of analog signals are acquired individually and processed jointly. Such setups are referred to as {\em distributed quantization}. In this work we propose a distributed quantization scheme for representing a set of sparse time sequences acquired using conventional scalar ADCs.
Our approach utilizes tools from \rev{secure} group testing theory to exploit the sparse nature of the acquired analog signals, obtaining a compact and accurate representation while operating in a distributed fashion.  We then show how our technique can be implemented when the quantized signals are transmitted over a multi-hop communication network providing a low-complexity network policy for routing and signal  recovery. Our numerical evaluations demonstrate that the proposed scheme  notably outperforms conventional methods based on the combination of quantization and compressed sensing tools.

\end{abstract}

\vspace{-0.4cm}
\section{Introduction}\label{intro}
\vspace{-0.1cm}
Physical signals typically have continuous-valued amplitudes. In order to process these signals using digital hardware, they \rev{are} quantized, namely, represented using a finite number of bits
\cite{gray1998quantization}. The conversion of an analog signal into a digital representation is  carried out using \acp{adc}, and consists of two steps: The \rev{signal} is first sampled \rev{in time, resulting in a discrete time sequence which is then}  quantized, \rev{often} by applying an identical uniform mapping to each  sample, i.e., uniform scalar quantization~\cite{eldar2015sampling}.

Conventional quantization theory considers the acquisition of a \rev{discrete time} analog source into a digital form \cite{gray1998quantization}. In some practical applications, such as sensor networks, multiple signals are acquired in distinct physical locations, while their digital representation is utilized in some central processing device, resulting in a distributed quantization setup. The recovery of a single parameter from the acquired signals was considered in \cite{gubner1993distributed, lam1993design} and its extension to the recovery of a common source, known as the CEO problem, was studied in \cite{berger1996ceo, oohama1998rate}, see also \cite[Ch. 12]{el2011network}.  Joint recovery of sources acquired in a distributed manner was studied in \cite{shlezinger2019joint}, which focused on sampling, while \cite{saxena2006efficient, wernersson2009distributed} proposed non-uniform quantization mappings for the representation of multiple sources\rev{. M}ultivariate (vector) quantizers for arbitrary networks were considered in \cite{fleming2004network}.

When restricted to using uniform \acp{adc}, the accuracy of the resulting digital representation is limited, depending on the number of bits utilized \cite[Ch. 23]{polyanskiy2014lecture}. It was recently shown that the effect of this quantization error can be significantly reduced by accounting for a specific task \cite{shlezinger2019hardware, shlezinger2018asymptotic,salamatian2019task}, or  the presence of a signal structure, as in \cite{cohen2019serial}, which considered  scalar quantization of a  sparse signal.

\begin{figure}
	\centering
	{\includegraphics[trim=0cm 0.4cm 0cm 0cm, width = \columnwidth]{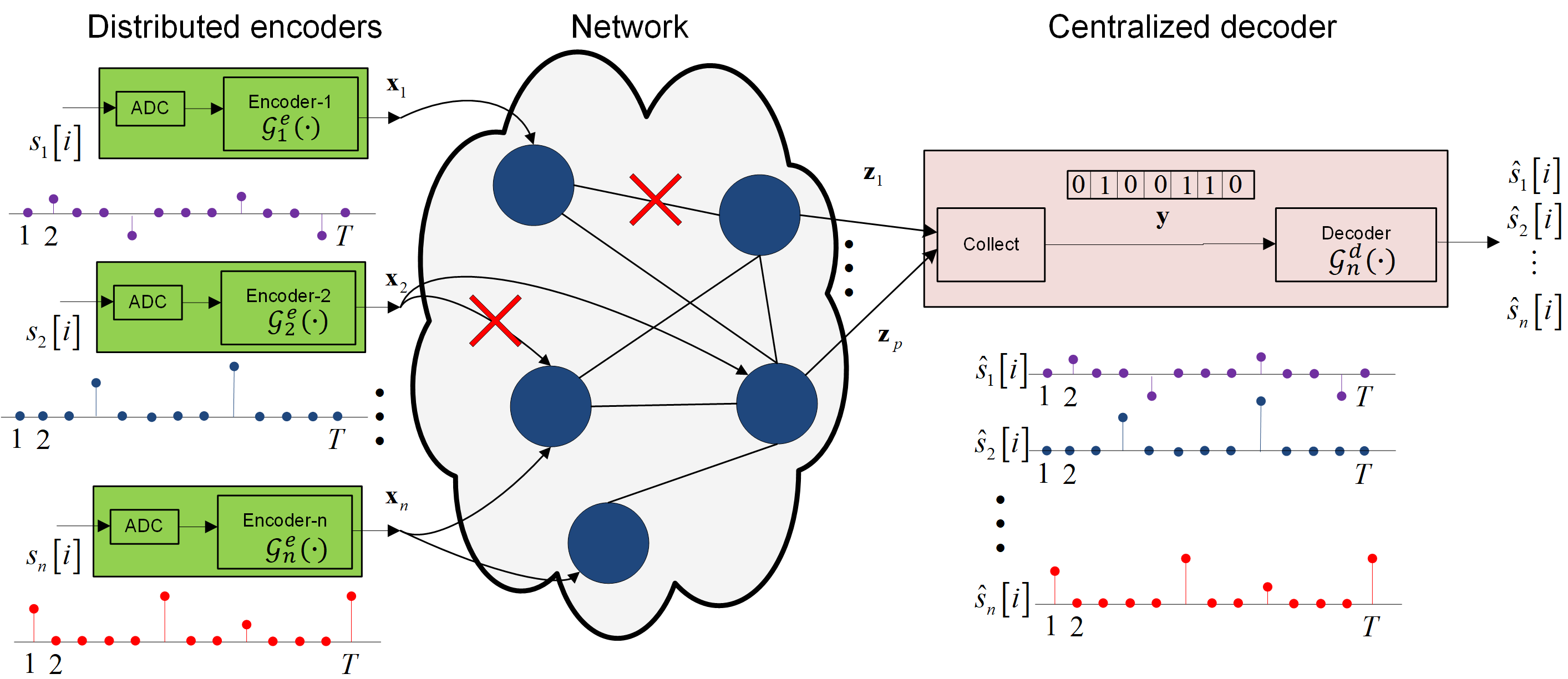}}
	\caption{Distributed quantization system illustration.}
	\vspace{-0.6cm}
	\label{fig:QCS_System}
\end{figure}

Sparse signals are encountered in a broad range of applications. \ac{cs} studies reconstruction of sparse signals from lower dimensional projections  \cite{eldar2012compressed}. Distributed \ac{cs} was studied in \cite{sarvotham2005distributed,baron2009distributed,do2009distributed,patterson2014distributed,feizi2010compressive}, sparse recovery from quantized projections was considered in \cite{jacques2013robust, boufounos20081,jacques2011dequantizing, gunturk2010sigma, kipnis2018single,boufounos2015quantization,saab2018quantization}, while \cite{shirazinia2014distributed,leinonen2018distributed} proposed vector quantization schemes for bit-constrained distributed \ac{cs}. Despite the similarity, there is a fundamental difference between distributed quantization of sparse signals and distributed \ac{cs} with quantized observations: In the quantization framework, the measurements are the sparse signals, while in \ac{cs} the observations are  a linear projection of the signals. Consequently, to utilize \ac{cs} methods in distributed quantization, one must first have access to the complete signal in order to project it and then quantize, imposing a major drawback when acquiring time sequences. This motivates the study of distributed quantization schemes for sparse time sequences, which is the focus here. \rev{A distributed quantization system is illustrated in Fig.~\ref{fig:QCS_System}}.

In this work we propose a distributed quantization scheme for jointly representing a set of individually observed \rev{jointly sparse sampled} time sequences. Such setups consist of a set of sensors, each observing a sparse sequence \rev{in discrete time}, and conveying its quantized observations to a centralized unit over a communication network, where it is used to formulate a digital representation of the observed signals.
Our scheme is specifically designed to utilize scalar uniform \acp{adc}, building upon our previous \rev{work on sequential quantization of sparse signals}  \cite{cohen2019serial}. In particular, \rev{we show how the quantization system of \cite{cohen2019serial}, which utilized tools from secure group testing theory \cite{cohen2016secure} to exploit sparsity in quantization, can be applied for distributed acquisition. Under a temporal joint-sparse model \cite[Sec. 3.2]{baron2009distributed}, the proposed method  achieves guaranteed accurate recovery while requiring a small number of bits for the overall representation. The resulting coding scheme, which operates over the binary field, allows improved reconstruction compared \ac{cs}-based methods which project the real valued observations prior to quantization.}

We first consider the case where each acquired signal is conveyed to the central unit via a direct link, representing, e.g., single-hop networks.
We characterize the achievable distortion of the \rev{proposed scheme} in the large signal size regime, showing that a given distortion level can be achieved with an overall number of bits which  grows logarithmically in the \rev{number of samples}.
 Then, we show how the technique can be extended to multi-hop networks, in which the quantized data  must travel over multiple intermediate links to reach the central server. We formulate simplified network policies, dictating the behavior of each intermediate node, and prove that the performance characterization derived for single-hop networks also holds here, as long as there exists at least a single path to the central unit. Our numerical results demonstrate that the proposed scheme achieves substantially more accurate digital representations compared to combining utilizing distributed-quantized \ac{cs} methods.

The rest of this paper is organized as follows:
Section~\ref{sys} introduces the system model.
Section~\ref{CodeConstruction} details the proposed distributed quantization scheme, while Section~\ref{simulation} provides simulation examples.

Throughout this paper, we use boldface lower-case letters for vectors, e.g., ${\myVec{x}}$;
the $i$th element of ${\myVec{x}}$ is written as $({\myVec{x}})_i$.
Matrices are denoted with boldface upper-case letters,  e.g.,
$\myMat{M}$,  $(\myMat{M})_{i,j}$   denotes its $(i,j)$th element. 
Sets are denoted with calligraphic letters, e.g., $\mathcal{X}$.
We use $\myI_{n}$ to denote the $n \times n$ identity matrix\off{.
%
We use 
$\bigvee$ to denote the Boolean OR operator}, while
$\mySet{R}$  and $\mySet{N}$ are the sets of real numbers  and natural numbers, respectively.

\vspace{-0.2cm}
\section{System Model}\label{sys}
\vspace{-0.1cm}
We consider distributed acquisition and centralized reconstruction of $\lenX$ analog time sequences. The sequences, denoted $\{s_m[i]\}_{m=1}^{\lenX}$  are separately observed over the period $i \in \{1,\ldots,\lenT\} \triangleq \mySet{\lenT}$, representing, e.g., sources measured at distinct physical locations. The signals are jointly sparse with joint support size $\SpaSize \ll \lenX\lenT$ \cite{baron2009distributed}. We focus on two models for the joint sparse nature of $\{s_m[i]\}$:

\paragraph*{Overall sparsity} Here, the ensemble of all $\lenX$ signals over the observed duration is $\SpaSize$-sparse, namely, the set $\{s_m[i]\}_{m \in \lenXset, i\in\mySet{T}}$ contains at most $\SpaSize$ non-zero entries. This model, in which no structure assumed on the sparsity pattern of each signal, coincides with the general joint-sparse model of \cite{baron2009distributed} without a shared component.
\paragraph*{Structured sparsity} In the second model the signals are sparse in both time and space. Specifically, for each $m \in \lenXset$, the signal $\{s_m[i]\}_{ i\in\mySet{T}}$ is $k_t$-sparse, while for any $i\in \mySet{T}$, the set $\{s_m[i]\}_{m \in \lenXset}$ is $k_s$-sparse. This setup is a special case of overall sparsity with $\SpaSize = k_s k_t$ with an additional structure which can facilitate recovery.

Each time sequence $\{s_m[i]\}_{i\in\mySet{\lenT}}$ is encoded  into a $\bits$-bits codeword denoted $\myX_m\in \{0,1\}^{\bits}$.
The encoding stage is carried out in a distributed manner, namely, each codeword $\myX_m$ is determined only by its corresponding time sequence $\{s_m[i]\}_{i\in\mySet{\lenT}}$ and is not affected by the remaining sequences. The codewords $\{\myX_m\}$ are conveyed to a single centralized decoder over a network, possibly undergoing several links over multi-hop routes. We consider a binary network model, such that each link can be either broken or error-free. The centralized decoder maintains links with $\lenZ$ nodes. The $\lenZ$ network outputs, denoted $\{\myZ_m\}_{m=1}^{\lenZ}$, \rev{are} collected by the decoder into a $\bits$-bits vector $\myY\in\{0,1\}^{\bits}$, which is decoded into a digital representation of the acquired signals, denoted $\{\hat{s}_m[i]\}$, as illustrated in Fig. \ref{fig:QCS_System}.
The system performance  is measured by the \ac{mse}   $\sum_i \sum_m\mathbb{E}\left[ (s_m[i]  - \hat{s}_m[i] )^2\right]$ and the quantization rate $\Rate = \frac{\bits}{\lenX\lenT}$.

We focus on the representation of time sequences, where each sample of the sparse source is observed in a different time instance. In order to avoid the need to store samples in analog, we require the acquisition to be carried out using  serial \acp{adc}, typically utilized by \acp{dsp} \cite{eldar2015sampling}. Here, the $m$th encoder operates on each sample $s_m[i]$ independently,  updating a register of $\bits$ bits, whose value upon the encoding of $s_m[i]$ is denoted by $\myX_{m,i} $. Once the complete vector time sequence is acquired, the encoder conveys the digital codeword $\myX_m\triangleq \myX_{m,\lenT} $. Note that both the encoders as well as the centralized decoder use $\bits$ bits for digital representation. 	

Our goal is to propose a distributed quantization system based on the above model. In particular, the distributed quantization scheme detailed in the following section consists of an encoding method, applied by each encoder; a decoding mapping, utilized by the central decoder; and a network behavior guidelines, namely, how the codewords $\{\myX_m\}$ are routed over the network.

\vspace{-0.2cm}
\section{Distributed Quantization Scheme}\label{CodeConstruction}
\vspace{-0.1cm}
In this section we detail the proposed distributed quantization scheme. We first consider a single-hop network in Subsection \ref{subsec:Code_SingleHop},
and incorporate the presence of a multi-hop network in Subsection \ref{subsec:Code_MultiHop}. A theoretical performance analysis and a discussion are provided in Subsections \ref{subsec:Code_Performance}-\ref{subsec:Code_Discussion}, respectively.

\vspace{-0.3cm}
\subsection{Single-Hop Networks}
\label{subsec:Code_SingleHop}
\begin{figure}
	\centering
	{\includegraphics[trim=0cm 0.4cm 0cm 0cm, width = \columnwidth]{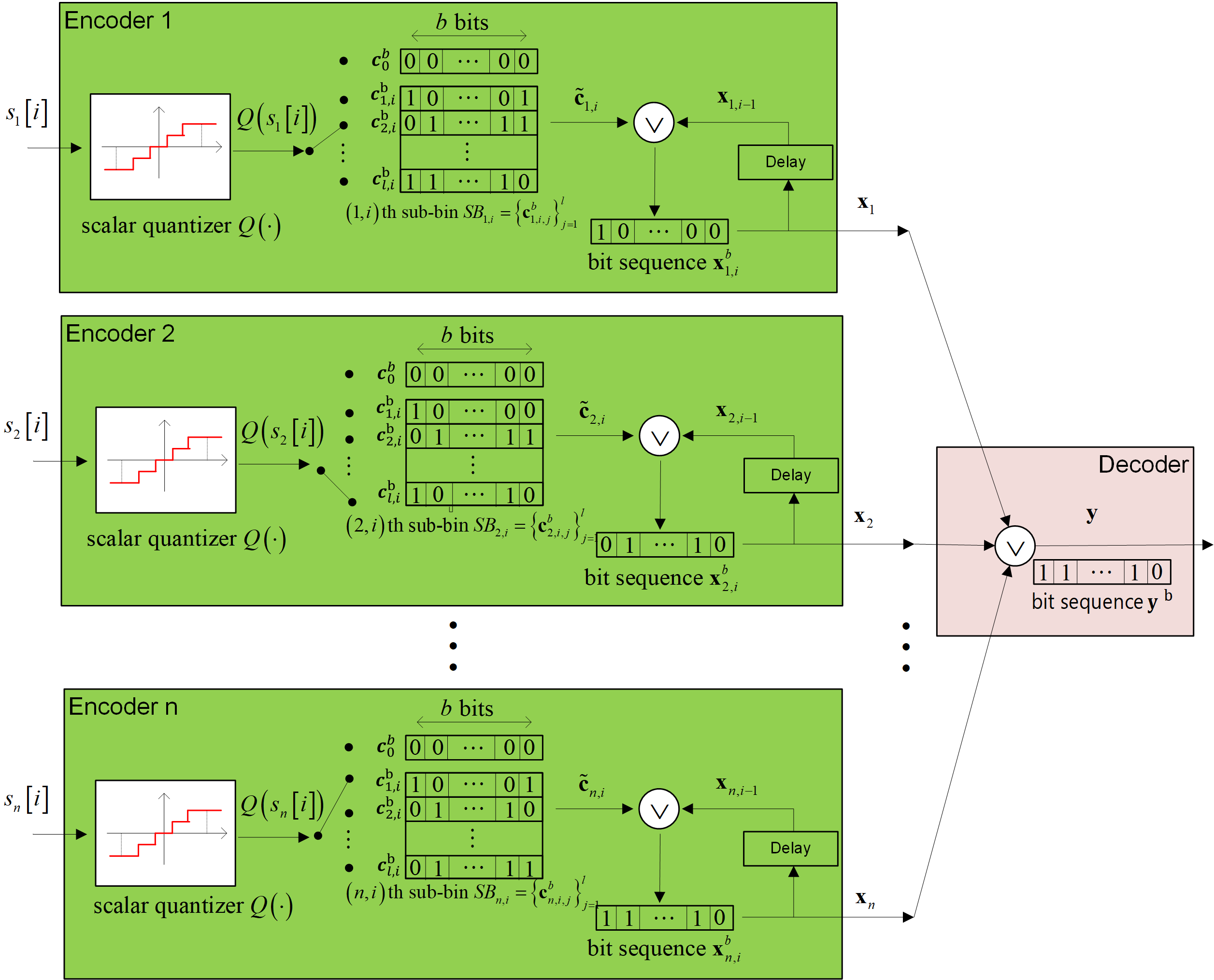}}
	\caption{Acquisition process in single-hop networks.}
	\vspace{-0.6cm}
	\label{fig:QCS_SingleHop}
\end{figure}

\vspace{-0.1cm}
In a single-hop network each encoder has a direct error-free link to the centralized decoder.
As mentioned above, we design our scheme to utilize serial scalar \acp{adc} to acquire each incoming sample, avoiding the need to store in analog previous samples required when using, e.g., \ac{cs}-based methods or vector quantization techniques. This is done in two steps. First, each encoder utilizes its own \ac{adc}, operating as a uniform scalar quantizer with resolution $\lenL+1$, to update a local register of $b$ bits. The relationship between $b$ and $\lenL$, as well as the remainig system parameters, are discussed in Subsection \ref{subsec:Code_Performance}. Once the acquisition of all $T$ time instances is complete, the encoders report  the binary vector stored in their local register to the central decoder over the single-hop network. The decoder then uses the received codewords to jointly produce a digital representation of all the signals. The acquisition pipeline is illustrated in Fig.~\ref{fig:QCS_SingleHop}.

\rev{The encoding procedure at each encoder is based on the  method proposed in our previous work   \cite{cohen2019serial}, which combined scalar \acp{adc} with group testing tools for serial quantization of sparse signals. The scheme of \cite{cohen2019serial} applied the same \ac{adc} to each incoming input, and assigned to each \ac{adc} output a codeword taken from a code bin determined by the time instance, which is in turn combined with the previous codewords using logical operations. We identify that the associative nature of logical operations allows this scheme to be carried out in a distributed manner. Here, instead of using a different code bin for each time instance, we use a different bin for each {\em user} for each time instance, i.e., we utilize sub-binning. Using this modification, the scheme proposed in \cite{cohen2019serial} can be applied for distributed quantization of jointly sparse signals, as detailed next. }

\subsubsection{Codebook Generation} \label{subsec:codebook}
Each encoder maintains a codebook which is known to the central decoder. These codebooks can be generated offline, either by the central decoder or by each remote encoder individually, following a random binning strategy.
Specifically, each codebook consists of $\lenL \cdot \lenT$ binary sequences of length $b$, drawn in an i.i.d. fashion from a Bernouilli distribution with parameter $\ln(2)/\SpaSize$. The $m$th encoder codebook is denoted by $\myBin_m$, $m \in \lenXset$.
The codebook is divided into $\lenT$ distinct subsets of equal size, referred to as {\em sub-bins}, denoted \rev{by} $\mySubBin_{m,i}\triangleq \{\myCodeword_{m,i,j}\}_{j=1}^{\lenL}$, $i \in \lenTset$. Finally, each codebook contains the zero codeword denoted by $\myCodeword_{0}$, such that $\myCodeword_{m,i,0} = \myCodeword_{0}$. The set of $\lenX$ codebooks thus contains  a total $\lenL \cdot  \lenX \cdot \lenT +1$ codewords.


\subsubsection{Encoder Structure} \label{subsec:encoder}	
As depicted in Fig. \ref{fig:QCS_SingleHop}, each incoming sample $s_m[i]$ is first quantized using a scalar \ac{adc} with resolution $\lenL + 1$, denoted  $Q_{\lenL}(\cdot)$, yielding a discrete value from the set $\{\ScaQuant_j\}_{j=0}^{\lenL}$, where  $Q_{\lenL}(0) = q_0$. The encoder uses the discrete value as index to select a codeword from its $i$th sub-bin, i.e., if $Q_{\lenL}(s_m[i]) = \ScaQuant_j$ then the codeword  $\myCodeword_{m,i,j} \in \mySubBin_{m,i}$ is chosen. Finally, the encoder updates a local $b$-bits register whose value at time instance $i$ is $\myX_{m,i}$ via
\vspace{-0.1cm}
\begin{equation}
    \myX_{m,i}  = \myX_{m,i-1}  \bigvee \myCodeword_{m,i,j},
   \label{eqn:RegUpdate1}
\vspace{-0.1cm}
\end{equation}
where \rev{$\bigvee$ is the Boolean OR operator, and} $\myX_{m,0}\equiv \myVec{0}$.
After the  sequence is acquired, $\myX_{m} = \myX_{m,\lenT}$ is conveyed  to the decoder.

\subsubsection{Decoder Structure} \label{subsec:decoder}
The decoder uses the received $\{\myZ_m\}_{m=1}^{\lenZ}$, which at the single hop case are given by $\myZ_m = \myX_m$ and $\lenZ = \lenX$, to recover the sparse signals.
To that aim, it first updates a single shared $b$-bits register $\myY$ based on the network outputs via
\vspace{-0.1cm}
\begin{equation}
\label{eqn:recovery}
    \myY  = \bigvee_{m=1}^{p}\myZ_m.
\vspace{-0.1cm}
\end{equation}
To obtain a digital representation of the $\lenX$ signals $\{\hat{s}_m[i]\}$ from  $\myY$, the decoder uses a \ac{ml} decoding scheme. To formulate the \ac{ml} rule, let $\vartheta$ be the number of possible sets of non-zero entries in the set of $\lenX$ signals. The value of $\vartheta$ depends on the nature of the joint-sparse signals. For example, for the general case of overall sparsity, $\vartheta = {\lenX\lenT \choose \SpaSize}$, while for structured sparsity  $\vartheta =  {\lenX \choose \SpaSize_t}{\lenT \choose \SpaSize_s}$. For other jointly sparse models, such as joint sparsity with a common component \cite[Ch. 3.2]{baron2009distributed}, different values of $\vartheta$ are used.
Let $\{\mySet{X}_w\}_{w \in \{1,\ldots, \vartheta\}}$ denote the possible support for the non-zero entries of vectorization of the time sequences, namely, for a given $w$, each element in $\mySet{X}_w$ is a pair $(m,i)$ indicating that $s_m[i] \neq 0$.
Following  \cite{cohen2019serial}, the decoder implements the following steps:
	\begin{itemize}
		\item For a given  $\myY$, the decoder recovers a collection of $\SpaSize$ codewords $\hat{\myMat{C}}_{\mySet{X}_w} = \{\myCodeword_{m,i,j_{m,i}}\}_{(m,i) \in \mySet{X}_w}$, {\em each one taken from a separate sub-bin}, for which $\myY$ is most likely, namely,
		%
		\begin{equation}
		\label{eqn:MLDef}
		\Pr\left(\myY  \big|  \hat{\myMat{C}}_{\mySet{X}_w} \right) \ge \Pr\left(\myY  \big|  \hat{\myMat{C}}_{\mySet{X}_{\tilde{w}}} \right), \quad \forall \tilde{w} \neq w.
		\end{equation}
		The decoder looks for both the set of $\SpaSize$ sub-bins $\mySet{X}_w$ as well as the selection of the codeword for each sub-bin, i.e., the selection of codeword index $j_{m,i}$ within $\mySubBin_{m,i}$, $(m,i) \in \mySet{X}_w$, which maximize the conditional probability \eqref{eqn:MLDef}.
		\item The decoder recovers $\{\hat{\myS}_{m}[i]\}$ from  $\hat{\myMat{C}}_{\mySet{X}_w}$ by setting its $(m,i)$th entry, denoted $\hat{s}_m[i]$, to be $\hat{s}_m[i] =\ScaQuant_{j_{m,i}} $ for  each $(m,i) \in \mySet{X}_w$ and  $\hat{s}_m[i] = \ScaQuant_0$ for $(m,i) \notin \mySet{X}_w$.
	\end{itemize}
	
The main rationale of the proposed scheme is that it generates the codebooks such that the codewords utilized by each encoder at each time instance, which are determined by the quantized values $\{Q_{\lenL}(s_m[i])\}$, can be recovered from $\myY$ with high probability. This property, which is discussed in Subsection \ref{subsec:Code_Performance}, stems from the fact that the coding scheme is in fact based on group testing tools, and particularly, on secured group testing \cite{cohen2016secure}. \rev{Note that the division of the codewords into per-user sub-bins allows the decoder to reduce the possible sets of codewords resulting in $\myY$, thus decreasing the computational burden compared to searching over the complete set of codewords.} In addition to its distributed nature, the proposed scheme can be applied over multi hop networks, as detailed in the sequel.

\vspace{-0.3cm}
\subsection{Multi-Hop Multipath Networks}
\label{subsec:Code_MultiHop}
\vspace{-0.1cm}
We now generalize our scheme to a multi-hop network, in which multiple directed links relate the distributed encoders and the centralized decoder. The intermediate nodes in the networks, which act as helpers or relays, can perform basic operations on their input from incoming links. For the sake of space and exposition, we consider a simplified model for this communication network, in which links are assumed to support $b$-bits of information without errors, or result in a complete erasure. We also assume that the transmission is synchronized, i.e., the encoders and intermediate nodes all transmit in sync across their outgoing links, and that the network is acyclic. Note that despite its simplicity, this model is reminiscent of several network models used in the literature, e.g., \cite{el2011network}. 

\begin{figure}
	\centering
	{\includegraphics[trim=0cm 0.4cm 0cm 0cm, width = 0.75\columnwidth]{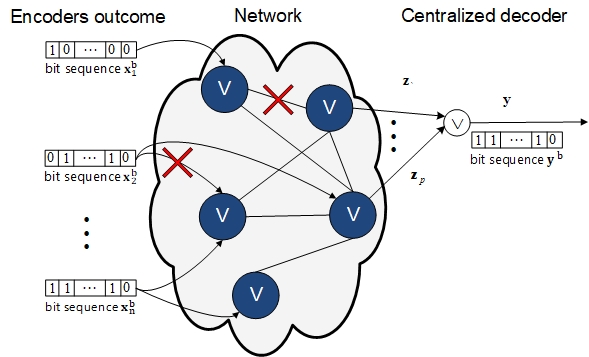}}
	\caption{Acquisition process in multi-hop multipath Networks.}
	\vspace{-0.6cm}
	\label{fig:QCS_MHP}
\end{figure}

The operation of the encoders and the decoder in the multi hop setup is identical to that discussed for single hop networks in Subsection \ref{subsec:Code_SingleHop}. The only addition is in the network policy, \rev{as depicted} in Fig. \ref{fig:QCS_MHP}: At each intermediate node, we perform a Boolean OR operation of all incoming input vectors (which is the same mathematical operation performed by the encoder and decoders in Subsection~\ref{subsec:Code_SingleHop}), and transmit the result length $b$-vector on all outgoing links. The network outputs are collected in $\myY$ via \eqref{eqn:recovery}.

Clearly, the resulting bit sequence at the decoder $\myY$ is identical to the one in  Subsection~\ref{subsec:Code_SingleHop}, as long as there exist at least one path in the network from each encoder to the centralized decoder. Note that this is in  contrast with the previous literature on distributed \ac{cs} over networks, where it is typical to impose  conditions on the network topology that guarantee a successful description \cite{feizi2010compressive}.
Consequently, the structures of the encoders and the decoder are invariant of whether the encoders communicate with the decoder directly or over multi-hop networks.
Additionally, the scheme we propose is robust to link failures:  as long as there exist at least one path from all encoders to the decoder, any number of link failures in the network still leads to the same received vector at the decoder, \rev{i.e., the coding scheme can achieve the min-cut max-flow bound of the network \cite{el2011network,dantzig2003max}.} 
The achievable performance of the scheme, whether applied over a single hop network or over multiple hops, is detailed in the following.

\vspace{-0.4cm}
\subsection{Performance Analysis}
\label{subsec:Code_Performance}
\vspace{-0.1cm}
Here, we analyze the achievable \ac{mse} of the proposed distributed quantization method.
As discussed in Subsection \ref{subsec:Code_SingleHop}, when the decoder successfully identifies the utilized codewords, the digital representation is $\hat{s}_m[i] = Q_{\lenL}(s_m[i])$, resulting in the \ac{mse}
\vspace{-0.1cm}
	\begin{equation}
	\label{eqn:ScaMSE2}
	    D(\lenL) \triangleq \frac{1}{\lenX\lenT}\sum\limits_{m=1}^{\lenX}\sum\limits_{i=1}^{\lenT} \mathbb{E}\left[ \left(s_{m}[i] - Q_{\lenL}(s_m[i])  \right)^2\right].
\vspace{-0.1cm}
	\end{equation}
Note that \eqref{eqn:ScaMSE2} is determined by the distribution of $\{s_m[i]\}$, the quantization mapping $Q_{\lenL}(\cdot)$, and its resolution $\lenL$. When $Q_{\lenL}(\cdot)$ represents a uniform mapping as in conventional \acp{adc}, $ D(\lenL)$ can be made arbitrarily small by increasing the internal parameter $\lenL$ \cite[Ch. 23]{polyanskiy2014lecture}. The effect of $\lenL$ on the quantization rate $\Rate$ for which the decoder can achieve $D(\lenL)$ with high probability is stated in the following theorem:
\begin{theorem}\label{direct theorem}
	The proposed distributed quantization scheme  achieves the average \ac{mse}  distortion $D(\lenL) $ in the limit  $\lenX\lenT \rightarrow \infty$ with $\SpaSize=\mathcal{O}(1)$  when the quantization rate $\Rate$ satisfies the following inequality:
\vspace{-0.1cm}
	\begin{eqnarray}\label{eq:reduce_hw}
	\Rate \ge \Rate_\varepsilon(\lenL)  \triangleq \max_{u \in\mySet{I}(\SpaSize)  }\frac{(1+\varepsilon)k}{u \cdot \lenX T}\log\left( \vartheta\cdot \lenL^u\right) ,
\vspace{-0.1cm}
	\end{eqnarray}
	for some $\varepsilon>0$.
	The set $\mySet{I}(\SpaSize)$ depends on the type of joint sparsity: for overall sparsity, $\mySet{I}(\SpaSize) = \{1,\ldots,\SpaSize\}$, while for structured sparsity $\mySet{I}(\SpaSize= k_s k_t) = \{u_t u_s: 1 \leq u_t \leq \SpaSize_t, 1 \leq u_s \leq \SpaSize_s\}$.
\end{theorem}

\begin{IEEEproof}
the proof follows similar arguments as in \cite[Appendix A]{cohen2019serial}, and is thus omitted for brevity.
\end{IEEEproof}

Theorem \ref{direct theorem} allows to determine what quantization rate $\Rate$ should be configured to achieve a desired quantization error. In particular, one should first set $\lenL$ to be the minimal value for which $D(l)$ is not larger than the desired error, and then set the quantization rate to be larger than $\Rate_\varepsilon(\lenL)$ for some small $\varepsilon$. Theorem \ref{direct theorem} then guarantees that, when $\lenX \lenT$ is sufficiently large, a digital representation of the desired accuracy is achieved with high probability.

\vspace{-0.3cm}
\subsection{Discussion}
\label{subsec:Code_Discussion}
\vspace{-0.1cm}
The proposed distributed quantization schemes has several practical advantages. First, it is designed to utilize conventional scalar \acp{adc}, carrying out acquisition in a serial manner, as opposed to \ac{cs}-based methods which require the complete time sequence to be available such that it can be projected and quantized.
In addition to this practical benefit, our proposed scheme also achieves improved performance compared to \ac{cs} schemes, as  illustrated in Section~\ref{simulation}.

Furthermore, our proposed method is  extendable for scenarios in which the remote encoders are connected to the centralized decoder via a multi-hop network. As discussed in Subsection \ref{subsec:Code_MultiHop}, the presence of such a network does not affect the system operation or its achievable performance, and only requires a simplified network policy to be carried out by the intermediate network nodes. While our analysis assumes that each encoder has at least a single path to the decoder, it can be shown that the presence of missing paths for some encoders does not affect the recovery of the remaining signals. In particular, by treating the output of a broken link as the zero vector, if the $m$th encoder has no path to the decoder, the recovery of $\{s_j[i]\}_{j\neq m}$ remains intact, while $\hat{s}_m[i]$ is estimated as being all zeros.
Finally, we note that while the quantization system  is specifically designed to exploit joint  sparsity to improve the recovery accuracy, it is also applicable with large $\SpaSize$.  though the complexity increases. In particular, it has been shown in the group testing literature that such coding schemes are capable of accurate decoding, which in our case implies an \ac{mse} of $D(\lenL)$, when $\SpaSize$ grows in the order of $o(\lenX\lenT)$ \cite{aldridge2017almost}. Furthermore, conventional group testing treats the encoding of binary data, while we consider that of $\lenL +1$ different values $\{q_j\}_{j=0}^{\lenL}$, hinting that larger values of $\SpaSize$ can be accurately recovered in the distributed quantization setup compared to standard group testing results. We leave the analysis of these conditions to future work.

\vspace{-0.2cm}
\section{Numerical Evaluations}\label{simulation}
\vspace{-0.1cm}
In this section we numerically evaluate the proposed distributed quantization method, compared to schemes based on distributed and quantized \ac{cs}.
To that aim, we consider a single hop network, and simulate $\lenX = 5$ time sequences. Each sequence consists of $\lenT = 20$ samples, following the overall sparsity model with support size $\SpaSize = 3$, where the non-zero indexes are generated uniformly, while their assigned values are randomized from an i.i.d. zero-mean unit variance Gaussian distribution.

\begin{figure}
	\centering
	{\includegraphics[trim=0cm 0.5cm 0cm 0cm, width = \columnwidth]{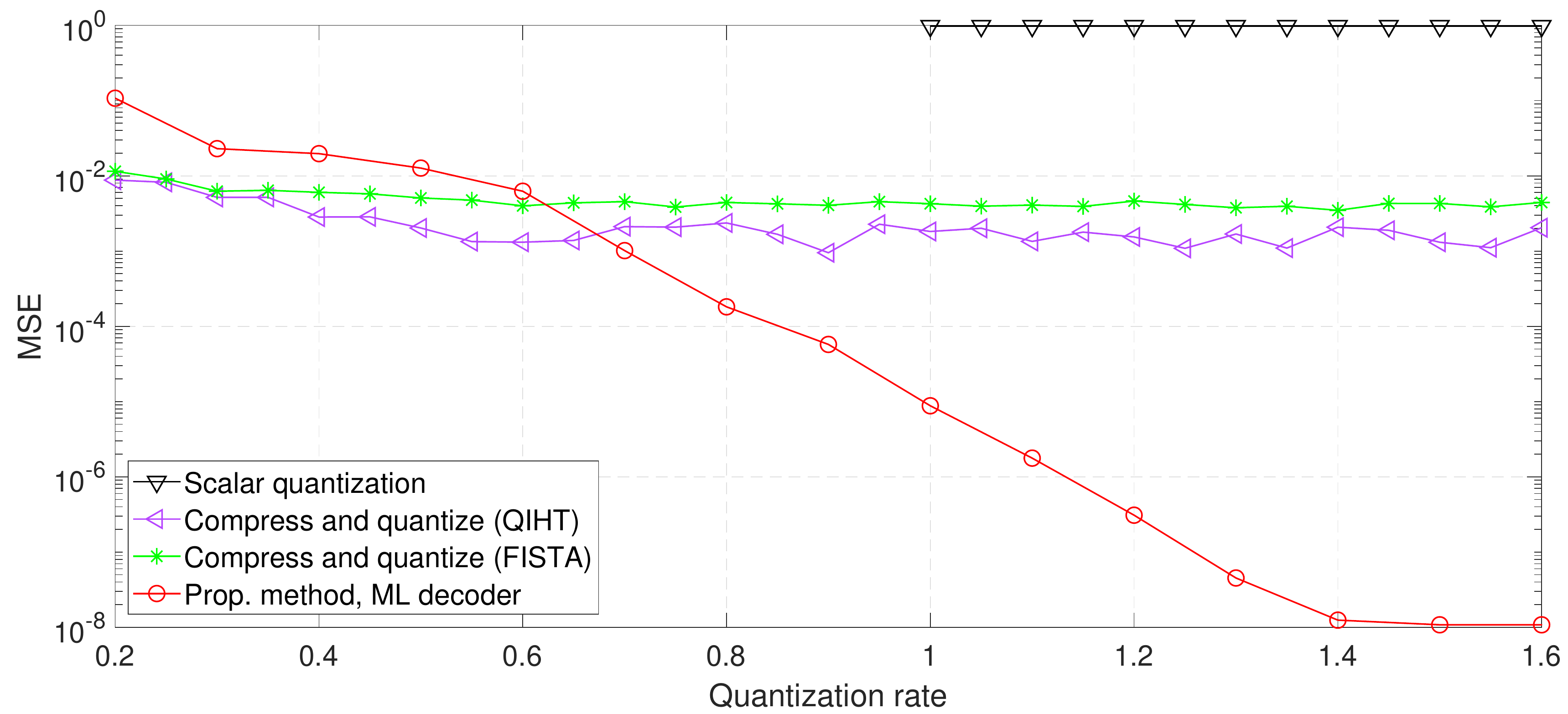}}
	\caption{Achievable distortion versus quantization rate $\Rate$.}
	\vspace{-0.4cm}
	\label{fig:distortion}
\end{figure}

In Fig. \ref{fig:distortion} we compare the \ac{mse} versus the quantization rate $\Rate$ achieved by our proposed  scheme to distributed  \ac{cs} methods with quantized observations. To guarantee that the used $\Rate$ satisfies \eqref{eq:reduce_hw}, we set $\lenL = \lfloor\frac{1}{\lenX\lenT}2^{\frac{\lenX\lenT\Rate}{\SpaSize(1+\varepsilon)}}\rfloor$, where   $\varepsilon$ is selected in the range $\varepsilon \in [0.8, 1.3]$. The \ac{adc} $Q_{\lenL}(\cdot)$ implements uniform quanization over $[-2,2]$. For distributed \ac{cs}, each signal $\{s_m[i]\}_{i\in\lenT}$ is compressed using i.i.d. zero-mean unit variance Gaussian projections into $\mySet{R}^{a}$, where the integer $a$ in the range $[\SpaSize,5\SpaSize]$ which minimizes the \ac{mse} is selected. Each set of projections is discretized using a uniform quanizer with support $[-2,2]$, where the resolution \rev{is selected, such that,} each encoder uses a total of $\Rate \cdot \lenT$ bits.
While more advanced schemes combining distributed \ac{cs} and vector quantization were proposed in \cite{leinonen2018distributed}, their complexity grows rapidly when $\lenX > 2$, and thus we focus on conventional distributed \ac{cs} with scalar quantization.
The quantized values are aggregated by the central decoder, which recovers the set of signals  using the \ac{qiht} method \cite{jacques2013quantized} as well as \ac{fista} \cite{beck2009fast}.   We also evaluate the case where each sample is separately uniformly quantized and  conveyed to the decoder without additional coding, modeling directly applying  scalar \acp{adc} for distributed acquisition.

Observing Fig. \ref{fig:distortion}, we note that the proposed distributed quantization scheme notably outperforms techniques based on distributed \ac{cs}. In particular, our method is shown to \rev{improve substantially} the accuracy of the overall digital representation as the quantization rate increases, while distributed quantized \ac{cs} is demonstrated to meet an error floor around $4\cdot 10^{-2}$ for \ac{fista} and $9\cdot 10^{-3}$ for \ac{qiht}. Standard uniform quantization, which is applicable only for $\Rate > 1$ as the \acp{adc} must utilize at least one bit, is notably outperformed by  the previous approaches, as it does not exploit the underlying sparsity.

\begin{figure}
	\centering
	{\includegraphics[trim=0cm 0.5cm 0cm 0cm, width = \columnwidth]{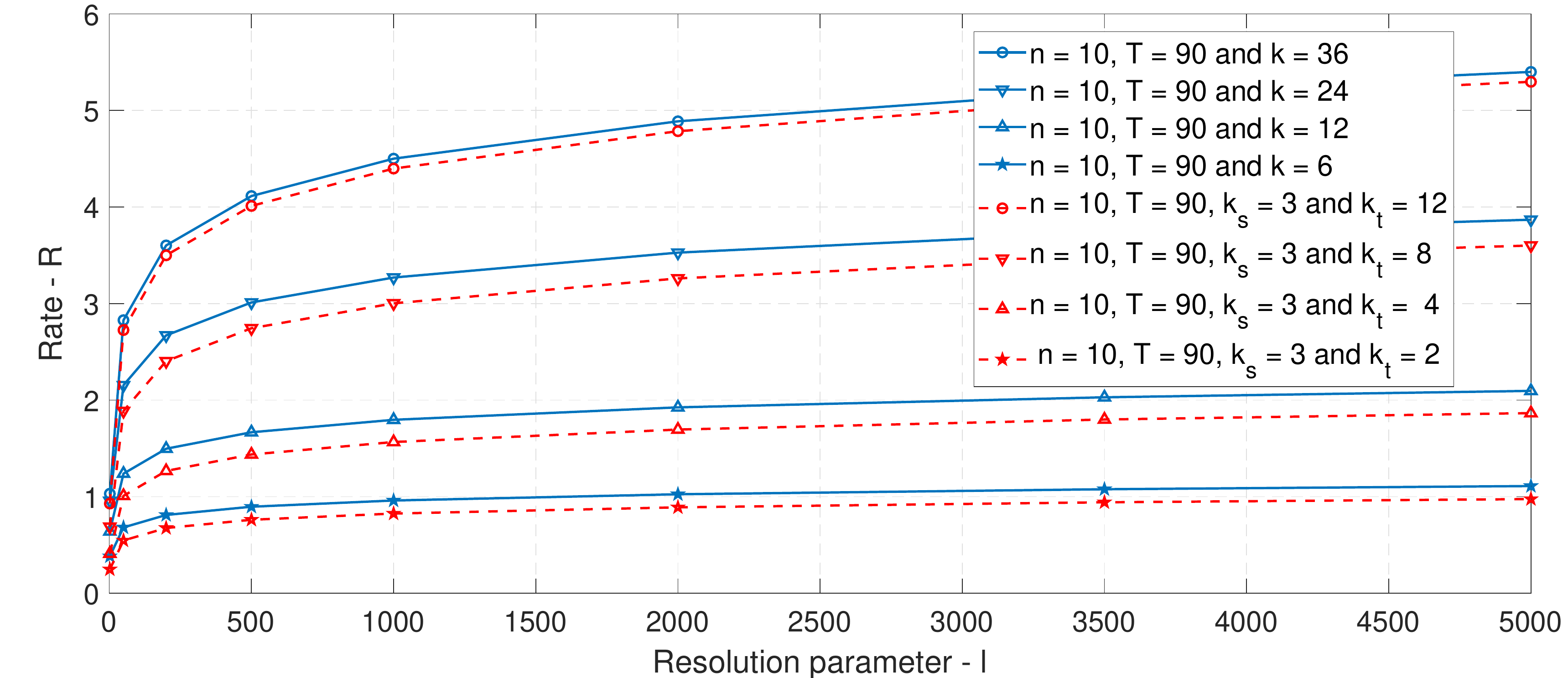}}
	\caption{Quantization rate threshold versus the resolution  $l$.}
	\vspace{-0.6cm}
	\label{fig:bound}
\end{figure}

In the study detailed in Fig. \ref{fig:distortion} we computed the achievable \ac{mse} for a given quantization rate. We note that Theorem \ref{direct theorem} allows us to \rev{determine rigorously} the quantization rate required to achieve a given \ac{mse}, as the latter is dictated by the quantization resolution $\lenL$. To demonstrate how the minimal quantization rate grows with the resolution $\lenL$, we compute in Fig. \ref{fig:bound} the minimal rate $\Rate_{\varepsilon}(\lenL)$ versus $\lenL$ for $\varepsilon = [0.8, 1.3]$. The setup evaluated here consists of $\lenX= 10$  sequence of $\lenT = 90$ samples each, for both overall sparsity with $\SpaSize \in \{6,12,24,36\}$ as well as structured sparsity with the same overall sparsity level and $k_s = 3$. Observing Fig. \ref{direct theorem}, we note that structured sparsity allows to use lower quantization rates, i.e., \rev{fewer} bits, to achieve the same level of distortion, due to the additional structure. We also note that the quantization rate grows slowly with $\lenL$, indicating that a minor increase in the quantization rate can allow the scheme to utilize \acp{adc} of much higher resolution, while maintaining the guaranteed performance of Theorem \ref{direct theorem}.

\vspace{-0.2cm}
\section{Conclusion} \label{conclusions}
\vspace{-0.1cm}
In this work we proposed a distributed quantization scheme designed to compactly and accurately represent a set  of sparse time sequences. Our proposed method utilizes serial scalar \acp{adc}, facilitating sequential acquisition while avoiding the need to store samples in analog, combined with coding schemes based on tools from group testing theory. We show how our approach can be naturally extended \rev{in} the presence of multi hop networks, by introducing simplified policies on the intermediate nodes, and derive sufficient conditions on the quantization rate required to achieve a desired quantization resolution. Our numerical study demonstrates that \rev{our} proposed method \rev{markedly} outperforms schemes based on distributed and quantized \ac{cs}, and illustrates how the presence of structured sparsity profiles can be exploited to utilize \rev{fewer} bits.
\newpage
\bibliographystyle{IEEEtran}
\bibliography{references,SecureNetworkCodingGossip}
\end{document}